\documentclass{nature}

\usepackage{amssymb}
\usepackage{graphicx}
\usepackage{epstopdf}
\usepackage[FIGTOPCAP]{subfigure}
\usepackage{color}
\usepackage[normalem]{ulem}
\usepackage{soul}
\usepackage[FIGTOPCAP]{subfigure}

\usepackage{floatrow}
\DeclareFloatFont{tiny}{\scriptsize}
\title{A kilonova from an ultra-quick merger of a neutron star binary}

\newcounter{firstbib}

\begin{document}

\maketitle
\author{Zhi-Ping Jin$^{1,2}$, Hao Zhou$^{1,2}$, Stefano Covino$^{3}$, Neng-Hui Liao$^{4}$, Xiang Li$^{1}$,  Lei Lei$^{1}$, Paolo D’Avanzo$^{3}$, Yi-Zhong~Fan$^{1,2}$,  and Da-Ming Wei$^{1,2}$.}
\begin{affiliations}
\small
\item{Key Laboratory of Dark Matter and Space Astronomy, Purple Mountain Observatory, Chinese Academy of Sciences, Nanjing 210023, China}
\item{School of Astronomy and Space Science, University of Science and Technology of China, Hefei 230026, China}
\item{INAF/Brera Astronomical Observatory, via Bianchi 46, I-23807 Merate (LC), Italy}
\item{Department of Physics and Astronomy, College of Physics, Guizhou University, Guiyang 550025, China}
\end{affiliations}

\hfill

\begin{abstract}
GRB 060505 was the first well-known nearby (at redshift 0.089) ``hybrid" gamma-ray burst that has a duration longer than 2 seconds but without the association of a supernova down to very stringent limits\cite{Fynbo2006}.  The prompt $\gamma-$ray flash lasting $\sim 4$ sec could consist of an intrinsic short burst 
and its tail emission\cite{Ofek2007}, 
but the sizable temporal lag ($\sim 0.35$ sec) as well as the environment properties led to the widely-accepted classification of  a long duration gamma-ray burst  originated from the collapse of a massive star\cite{McBreen2008,Thone2008,Thone2014}. Here for the {\it first} time we report the convincing evidence for a thermal-like optical radiation component in the spectral energy distribution of the early afterglow emission. In comparison to AT2017gfo, the thermal radiation is $\sim 2$ times brighter and the temperature is comparable at similar epochs.   
The optical decline is much quicker than that in X-rays,  which is also at odds with the fireball afterglow model\cite{Piran2004} but quite natural for the presence of a blue kilonova\cite{Li1998,Metzger2017}. Our finding reveals a neutron star merger origin of the hybrid GRB 060505 and strongly supports the theoretical speculation that some binary neutron stars can merge  ultra-quickly (within $\sim 1$ Myr) after their formation\cite{Belczynski2006} when the surrounding region is still highly star-forming and the metallicity remains low.
Gravitational wave and electromagnetic jointed observations are expected to confirm such scenarios in the near future.
\end{abstract}


Neutron star binaries, if close enough, are expected to merge with each other in the Hubble timescale after their formation\cite{Abadie2010}. Such mergers are able to give rise to  energetic gravitational wave radiation, short gamma-ray bursts if our line of sights are within the cone, and kilonova emission\cite{Eichler1989,Li1998,Metzger2017}, as measured in the remarkable/historic event of GW170817/GRB 170817A/AT2017gfo\cite{Abbott2017,Goldstein2017,Pian2017}. So far, GRB 170817A is still the unique short burst with a gravitational wave detection. For several previously-detected short GRBs,  the successful identification of kilonova signals in the afterglow data\cite{Tanvir2013,Berger2013,Jin2016,Troja2018,Jin2020} provides the strongest evidence for their neutron star merger origin. 
The merger timescales of neutron star binaries can be estimated from the properties of the population of stars in the host galaxy and a wide range of ${\cal O}(1)-10$ Gyr was reported (see Extended Data Table 1 for a summary).  Population synthesis studies however suggest the presence of a quick merger channel and the timescale could be as short as $\sim 10^{-3}-1$ Myr (the shortest delay time, including both the formation time of the compact object binary and the merger time, is $\sim 10$ Myr)\cite{Belczynski2006}, for which the surrounding region is still highly star-forming.  The successful identification of such ultra-quick merger event, which is the main goal of this work, would be essential for better understanding the formation and evolution of the compact binaries. 

\begin{figure}[!h]
\begin{center}
\includegraphics[width=0.99\columnwidth]{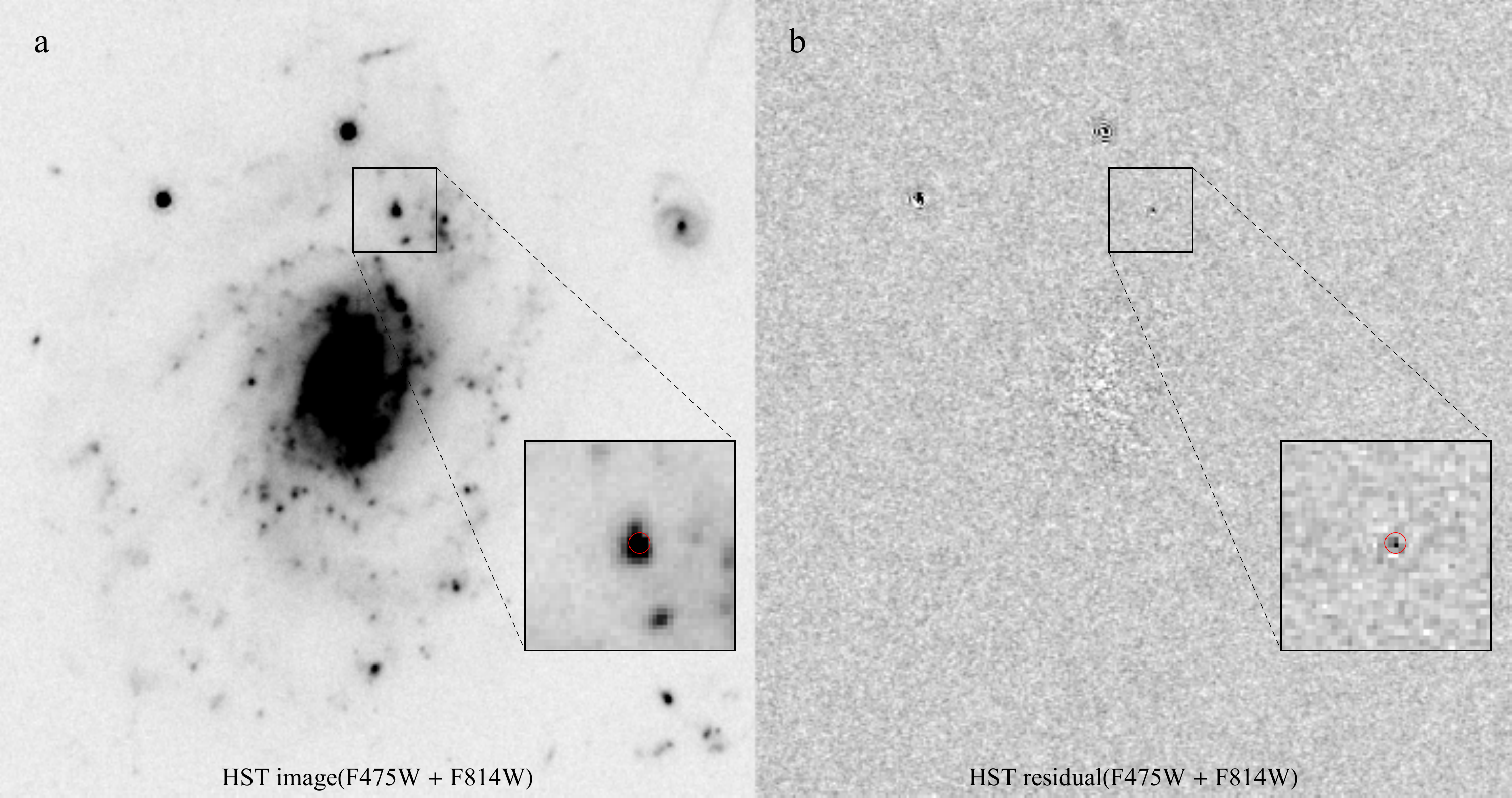}
\end{center}
\caption{
{\bf The HST observation and residual images.}
Panel (a) is the stacked HST F475W and F814W images. The open red circle marks the position of the afterglow revealed by VLT and Gemini observations\cite{Fynbo2006,Ofek2007,Xu2009}. Panel (b) is the stacked residual images of HST F475W and F814W observations between 15 May and 6 June 2006, which reveals a new source at exactly the same site of the early afterglow (i.e., the open red circle) reported in the literature.
}
\label{fig:Position}
\end{figure}

Though the so-called long GRBs with a duration longer than 2 seconds have been widely accepted to be originated from the collapse of the massive stars,  the nearby event GRB 060614 lasting about 102 s is an exception because of the lack of a bright supernova component in the late afterglow\cite{Fynbo2006}.  Both the compact merger origin and the failed supernova origin have been extensively discussed in the literature and this heavily debated event has been named as a hybrid GRB. A through re-examination of the plentiful data set of the hybrid GRB 060614 finally revealed the emergence of a very dim and soft optical component (i.e., the so-called red kilonova) at about 10 days after the burst, as expected in the neutron star$-$black hole merger scenario, favors the intrinsic ``short GRB" nature of this peculiar event\cite{Yang2015}.  GRB 060505 is the other well-known hybrid event with a duration of $\sim 4$ s. At variance without the plentiful multi-wavelength afterglow observations of GRB 060614, previously just upper limits on the optical emission of GRB 060505 were reported at $t>1.2$ days after the trigger \cite{Fynbo2006,Ofek2007} (see however Fig.\ref{fig:Position}b  for our late time detection), which renders the hunting for the GRB 060614-like red kilonova component impossible and leads to the conclusion that it is not suitable for the kilonova identification\cite{Jin2016}.  Moreover, for hybrid GRB 060614 the almost zero temporal lag as well as the  stellar population age of $\sim 0.57$ Gyr\cite{Leibler2010} are in agreement with being intrinsically ``short", for which the identification of a kilonova signal is  anticipated. While for hybrid GRB 060505,   though a ``short" nature was hypothesized before\cite{Ofek2007},  the sizable time lag as well as the very young stellar population were ``strongly" in favor of being a long burst\cite{McBreen2008,Thone2008,Thone2014}, for which the kilonova emission is unexpected. The detection of blue kilonova components at $t<1$ day after the trigger of a few short GRBs \cite{Pian2017,Jin2020}, anyhow, motivates us to carefully re-examine the optical/X-ray emission of hybrid GRB 060505.  {\it Surprisingly}, we find a distinct signal in the  spectrum, as reported below.

\begin{figure}[!h]
\begin{center}
\includegraphics[width=0.8 \columnwidth]{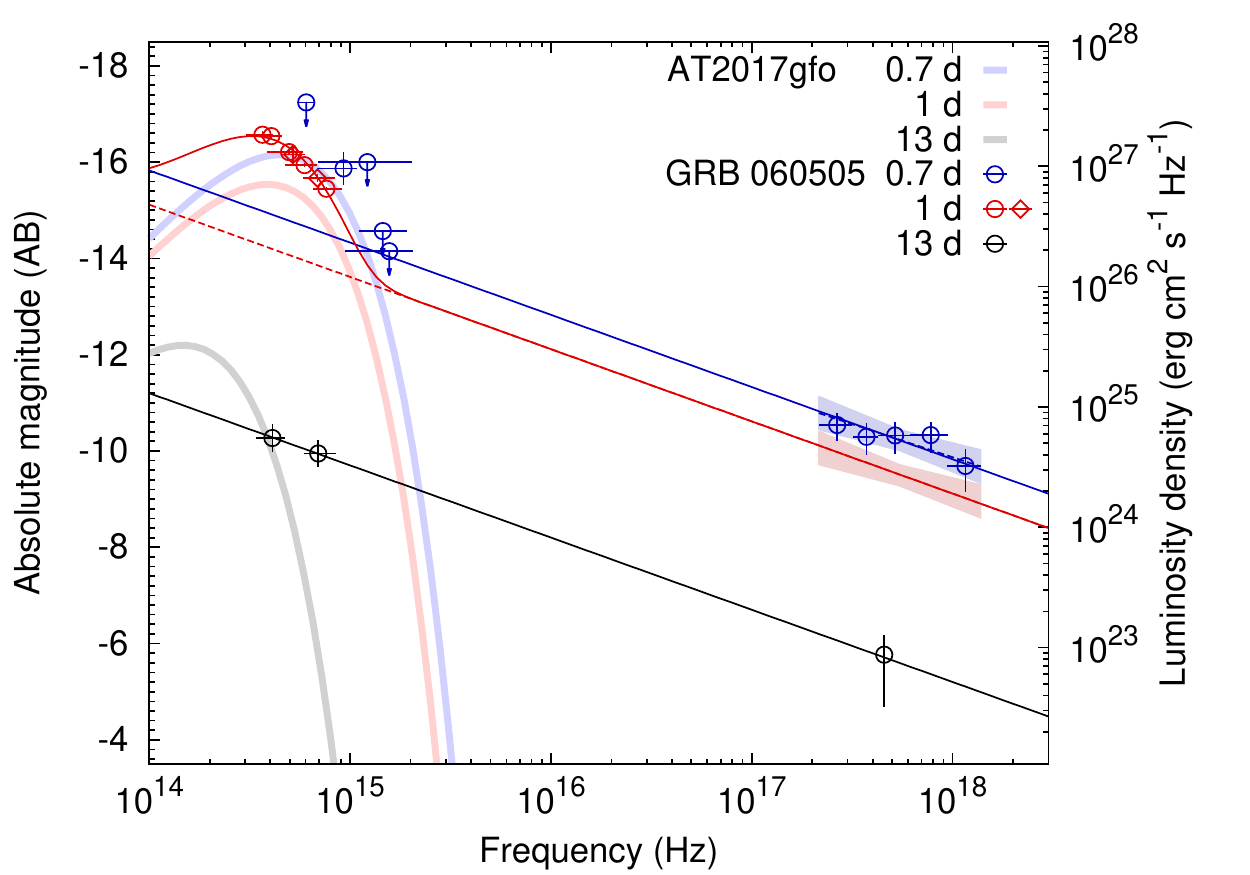}
\end{center}
\caption{
{\bf Spectral energy distributions  (SEDs) of the ``afterglow" emission of GRB 060505.}
Here we show the comparison of the optical and X-ray SED of the ``afterglow" radiation of GRB 060505 (i.e., the empty circle symbols) to the counterpart to GW170817 (i.e., the thick lines based on the temperatures reported in Drout et al.\cite{Drout2017}). Proper redshift corrections have been applied to both events. The optical measurements by VLT and Gemini at $t'=t/(1+z)\approx 1$ day are both presented. 
A single power-law spectrum of $\nu^{-0.6}$ is favored by the third epoch data as well as the UVW2 upper limit and the XRT  measurement in the first epoch. 
The second epoch data are fitted with a thermal component (i.e., $T=5750$ K)  plus a $\nu^{-0.6}$ power-law component. 
The U band detection in the first epoch is also significantly in excess of the power-law component, resembling the spectrum measured later at $t'=1$ day.   }
\label{fig:GRB060505}
\end{figure}

At variance with most {\it Swift} GRBs,  the on-board detection significance for GRB 060505 was below the threshold for an autonomous spacecraft maneuver. It was not recognized until 0.6 day after the trigger of the burst when the analysis of the full data set on the ground showed the burst to be statistically significant\cite{Palmer2006}.  Fortunately, the subsequent X-ray Telescope (XRT) observations started immediately and the detection in X-rays provided an accurate localization.  Consequently, {\it Swift}/UVOT, Very Large Telescope (VLT), Gemini, Hubble space Telescope (HST) and some other optical telescopes carried out follow-up observations\cite{Ofek2007,Xu2009}.  The X-ray emission was also detected at late times by XRT and Chandra\cite{Ofek2007}. In this work, we re-analyze all the observation data and search for the possible kilonova signal in the spectrum. In addition to confirming the previous optical measurements, we found that  GRB 060505 was also marginally detected by UVOT in U band with a signal to noise ratio (SNR) of $3.3$, and by HST in F475W and F814W bands (The total ${\rm SNR}$ is 5.9 for these two bands).  We have also analyzed the VLT $z$-band data, which was ignored before\cite{Xu2009}, and obtained a very significant detection (see the Methods for the details). The detailed analysis results are summarized in Extended Data Table 2, where the Galactic dust extinctions have been corrected.   
In Fig.\ref{fig:GRB060505} we present the time-resolved optical to X-ray spectra. We do not consider the internal redding of the spectrum since the high-resolution spatially resolved spectroscopy of the host galaxy  suggests a low extinction at the site of GRB 060505\cite{Fynbo2006,Thone2014}  (Our conclusions are unchanged if there was a moderate intrinsic redding, as demonstrated in the Methods). 
At $t\approx 14$ day, the extracted X-ray data and the HST optical data can be reasonably fitted by a single power-law spectrum of  
$F_\nu \propto \nu^{-0.58\pm0.09}$.  At $t\approx 0.7$ day, the {\it Swift} UVM2/UVW2 upper limits and the XRT data tightly constrain the spectrum to be harder than $\propto \nu^{-0.6}$, nicely in agreement with the late time HST/X-ray data. Interestingly, the early time {\it Swift}/XRT data also  favor a hard spectrum of $\propto \nu^{-0.55\pm 0.34}$. 
We therefore take an intrinsic optical/X-ray afterglow spectrum of $f_\nu \propto \nu^{-0.6}$, which is typical for the GRB afterglow and suggests that the typical synchrotron radiation frequency of the forward shock accelerated electrons is below the optical band while the cooling frequency is above the X-ray band\cite{Piran2004}. However, at $t\sim 1.1$ days after the trigger, the VLT and Gemini data can be fitted by a spectrum of  $\propto \nu^{-1.52\pm0.07}$, which is in agreement with the spectrum of $\propto \nu^{-1.7\pm 0.7}$  based on the $g'$ and $r'$  band Gemini data only\cite{Ofek2007} but much steeper than the intrinsic spectrum $\propto \nu^{-0.6}$ suggested by the early 
UVOT/XRT and late HST/Chandra data. Such a puzzle can be naturally solved if the VLT/Gemini optical fluxes were dominated by a thermal-like component. Indeed, we found that the VLT/Gemini optical data and the X-ray data can be fitted by a superposition of the thermal emission with an intrinsic (i.e., redshift corrected) temperature of $T_{\rm int}\approx 5750\pm110$ K ($\chi^2/{\rm d.o.f}\sim 2$) and a power-law component of $\propto \nu^{-0.6}$.  As for the UVOT observations at $t=0.71$ day since the trigger, though the emission was solely detected in U band, the tight constraints in other bands suggest an unusually soft power-law spectrum $\propto \nu^{-2.8}$, which is much steeper than the VLT/Gemini optical data and that inferred for the late time  HST/Chandra data.  The U band emission is also well above the XRT/UVW2 constraint. Very likely the UVOT data hosts a thermal spectrum with a temperature of $6300~{\rm K}<T_{\rm int}<9800~{\rm K}$. This temperature, though not uniquely determined, is higher than that found at $t=1.1$ day after the burst. Such a temperature decrease is natural for the kilonova emission, as observed in AT2017gfo\cite{Drout2017}.  

Besides the plentiful spectral evidence, the presence of an optical kilonova is also supported by the temporal behaviors of the optical and X-ray emissions (see Extended Data Fig. 1).  With the VLT/Gemini/HST data and the simplest power-law decline behavior, the optical decline is as steep as $t^{-2.26}$ in the $I$/F814W band and could be attributed to the post-jet break radiation of a narrowly beamed outflow\cite{Piran2004}. While the X-ray emission dropped with time as $t^{-1.39^{+0.24}_{-0.18}}$ {(a similar result was reported  in the literature\cite{Ofek2007})}, which is much more shallowly than the optical emission, in tension with the forward shock afterglow model\cite{Piran2004}. This tension can be solved supposing the $t<1.1$ day optical emission was dominated by a blue kilonova. Moreover, the U-band emission decline can be evaluated though indirectly. 
With the VLT/Gemini data at $t\approx 1.1$ day after the burst, the AB magnitude of U-band emission can be estimated to be ${23.49\pm0.16}$ 
assuming a thermal spectrum (the assumption of a power-law spectrum of $\propto \nu^{-2.8}$ would yield a similar result),
which is much dimmer than that measured at $t \approx 0.7$ day after the burst. The corresponding U-band decline is $t^{-2.7}$,  which is at odds with the forward shock emission but well consistent with the ultraviolet decline behaviors of the blue kilonova component of AT2017gfo\cite{Drout2017,Kasliwal2017}. As shown in Fig.\ref{fig:GRB060505}, in comparison to AT2017gfo, the thermal radiation is brighter by a factor of $\sim 2$ and the temperature is comparable at $t'=t/(1+z)\sim 0.6-1.0$ day (see also Extended Data Fig.2).   The X-ray emission of GRB 060505 is relatively faint (see Extended Data Fig.3), with which the forward shock optical emission is weak and hence the ``additional component" is easier to identify.
In view of the above various evidence/facts, it is robust to conclude the presence of a blue kilonova component in the early optical data of GRB 060505. 

With the inferred  temperature ($T_{\rm int}\approx 5750$K) and luminosity ($L\approx 8.35\times10^{41}$erg), the parameters of the hot outflow can be evaluated.  The bulk Lorentz factor $\Gamma=1/\sqrt{1-\beta_{\Gamma}^{2}}$  of the outflow is governed by  $(1+\beta_{\Gamma})\beta_{\Gamma}\Gamma \approx 0.4L_{42}^{1/2}(T_{\rm int}/6000~{\rm K})^{-2}(t'/1~{\rm d})^{-1}$, where $\beta_{\Gamma}$ is the velocity in unit of the speed of light ($c$). Therefore we have $\beta_{\Gamma}\approx 0.35$, which is in agreement with the kilonova model but disfavors the relativistic shock-breakout/cocoon\cite{Kasliwal2017} origin of this thermal radiation component. Supposing the observation time is comparable to the diffusion timescale of $(1+z)(\kappa M_{\rm ej}/4\pi \beta_{\Gamma}c^{2})^{1/2}$,  the mass of the outflow can be estimated as $M_{\rm ej}\sim 0.014~{M_\odot}~(\beta_{\Gamma}/0.35)(t'/1~{\rm d})^{2}(\kappa/1~{\rm cm^2~g^{-1}})^{-1}$, where $\kappa$ is the opacity parameter of the sub-relativistic lanthanide-poor outflow. This suggests an efficient mass ejection via the disk wind, similar to the case of AT2017gfo\cite{Pian2017}.  Interestingly, the blue kilonova in GRB 060505 is brighter than that of GW170817, but there is no clear evidence for the presence of a luminous red kilonova in the HST data of GRB 060505 (see Fig.\ref{fig:GRB060505}), indicating that the blue and red kilonova material were launched via different physical processes and they may be not tightly correlated. 

\begin{figure}[!h]
\begin{center}
\includegraphics[width=0.9 \columnwidth]{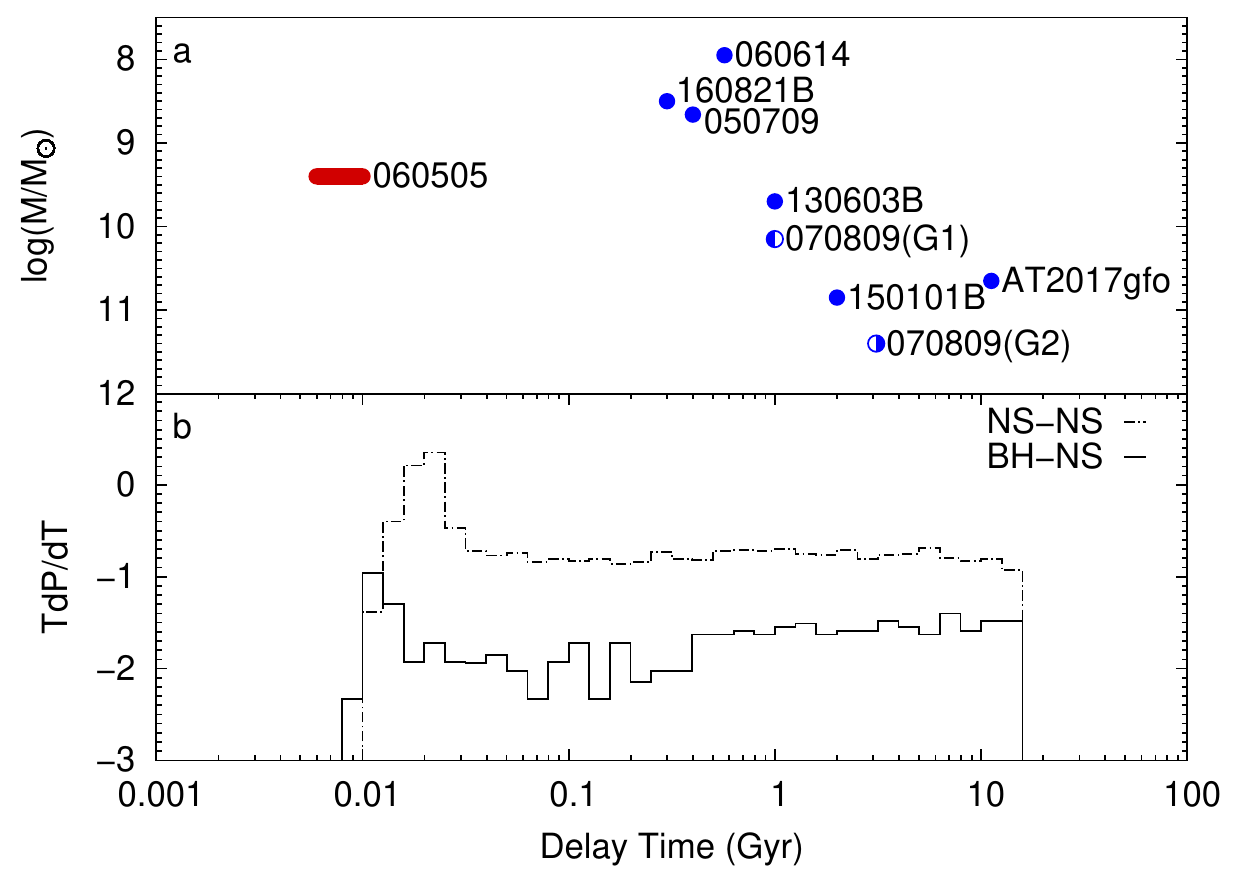}
\end{center}
\caption{{\bf The delay times of the compact object binaries giving rise to the kilonova-associated GRBs.} The delay time includes both formation time of the neutron star binary and its merger time. The histograms in the bottom pannel represent the predictions of time delay distributions of binary neutron star mergers or the neutron star$-$black hole mergers\cite{Belczynski2006}. Apparently, GRB060505 is typical in the host galaxy luminosities (the upper pannel) but has the shortest time delay\cite{Ofek2007,Thone2008,Leibler2010} detected so far (it is shorter than others by a factor of $\sim 10$s$-10^{3}$; see Extended Data Table 1 for the details), which is at the low end of the theoretical distribution, pointing towards an ultra-quick merger origin. }
\label{fig:Delaytime}
\end{figure}

At a  low redshift $z=0.089$, the site of GRB 060505 and its host galaxy has been observed/studied with high resolution in great detail, which is currently not possible for the majority of GRB hosts.  Besides being a hybrid event,  GRB 060505 is characterized by its spatial distribution in its host galaxy. As revealed by the optical observations, the GRB site is at the edge of an “island’’ with a diameter of $\sim 400$ pc ``separated" from the main part of the host galaxy\cite{Ofek2007} (see also panel (a) of Fig.\ref{fig:Position}) . 
The resolved data show that the GRB site has the lowest metallicity in the galaxy with $\sim Z_\odot/3$, but a typical specific star formation rate  of $\sim 7.4 M_\odot ~{\rm yr^{-1}}(L/L_{*})^{-1}$  in the host \cite{Thone2014}, where $L$ and $L_{*}$ is the luminosity and characteristic luminosity, respectively.  By fitting single-age spectral synthesis models to the stellar continuum, Th\"{o}ne et al.\cite{Thone2008} derived a very young age for the GRB site, supported by photometric and H${\alpha}$ line measurements, of $\approx 6\pm 1$ Myr (independent suggestions for a very young age, i.e., $\leq 10$ Myr, are available in the literature\cite{Ofek2007,Leibler2010}). 
This young age corresponds to the lifetime of a $\sim 30 M_\odot$ star and sets an stringent upper limit on the timescale of the progenitor star(s). Previously, these properties have been taken as the compelling evidence for a collapsar origin of GRB 060505\cite{Thone2008,Thone2014}. However, in this work we have found convincing evidence for the kilonova emission and GRB 060505 should have a compact object merger origin. Therefore, the neutron star binary system should merge within a time of a few Myr, which is  shorter than that of other short GRBs and hybrid GRB 060614 by a factor of tens or larger (see Fig.\ref{fig:Delaytime}) and consistent with  the shortest time delay expected for the merging compact object binary systems\cite{Belczynski2006}. To our knowledge, this is the first time to find the evidence for the ultra-quick merger channel of neutron star binaries in the GRB/kilonova data. There are two intriguing implications. One is that the neutron star binary system should be narrowly separated (i.e., $< 0.5 R_\odot$ if the system solely consisted of neutron stars) or have a high eccentricity. The other is that the death of the very massive stars (i.e., with an initial mass $\geq 30M_\odot$, supposing the delay time is as short as $\approx 6$ Myr) can indeed produce neutron star remnants, supporting the theoretical prediction of the stellar evolution\cite{Sukhbold2016}.  { Moreover, a sizable spectral lag, though very unusual for short events, can not be taken as a decisive indicator of long bursts any longer. Indeed, except GRB 060505 examined here,  GRB 170817A, the short burst associated with GW170817, has a spectral lag of $\sim 0.15$ s\cite{Goldstein2017} and a small fraction of BATSE short GRBs have a long spectral lag $\geq 0.1$ s\cite{Yi2006}.}

The local neutron star merger rate is constrained to be $320^{+490}_{-240}~{\rm Gpc^{-3}~yr^{-1}}$  (see Ref.\cite{Abbott2020a}) and the inclination angle averaged detection distance of the binary neutron star mergers in the full sensitivity run of the advanced LIGO/Virgo/KAGRA is $\approx 330$ Mpc\cite{Abadie2010}.  The detection rate of the binary neutron star mergers is thus $\approx 48^{+73}_{-32}~{\rm yr^{-1}}$. For most of these events, no associated GRBs will be observed due to the relativistic beam effect and the electromagnetic counterparts would rely on the survey measurements of the Legacy Survey of Space and Time and other telescopes. Please note that the presence of a GRB/afterglow provides a reliable estimate of the time, sky location, and distance of the binary merger signal, which significantly reduces the parameter space of a follow-up gravitational wave search and consequently enhances the detection distance by a factor of $\sim 2$  (see Ref.\cite{Kochanek1993}). Indeed, GRB 060505 with a low redshift of $z=0.089$ was already  within the horizon of advanced gravitational wave detectors if one takes into account such a sensitivity distance enhancement\cite{LiX2016}.  As long as the electromagnetic counterparts have been reliably identified, the dedicated spatially/spectrally resolved observation of the sites of these ``nearby" merger events would reveal the physical properties of the surrounding environment, with which our understanding of the stellar evolution is expected to be  significantly advanced.

\newpage

{\bf Methods}

{\bf The optical data analysis of GRB 060505.} Our optical data analysis results are reported in Extended Data Table 2. Below we introduce some key information of the analysis. 

{\it VLT data processing.} The ESO VLT observed the position of GRB 060505 on May 6 (under programme ID 077.D-0661, PI: Vreeswijk, P. M.). We have downloaded the archived VLT raw data together with necessary calibration and reference frames including bias, flat and standard stars. 
Standard recipes (bias subtraction, flat-field normalization) were applied for data reduction. 
Since the images in $z$ band suffered from strong fringing, we downloaded some other science images taken in the same observation method within half year, 
derived the fringing pattern and subtracted it. 
The object images taken on 6 May are subtracted by the host galaxy images taken on 14 September to reveal the afterglow. 
The afterglow is clearly detected in all $B$, $V$, $R$, $I$ and $z$ bands. 
Standard star field ``MARK A" is observed following the host galaxy observation on 14 September. These frames are used to determine zero point.
Since the $z$ band is not included in the Landolt standard star catalog, we have transformed it from the Johnson-Morgan-Cousins $UBVR_{C}I_{C}$ system using the relations proposed in Smith et al.\cite{Smith2002}.
The standard star field ``SA95-105" was observed in $z$ band 7 hours later, zeropoint can be derived by calibrating to the SDSS catalog, it is $26.47\pm0.04$th AB magnitude, which confirmed the transformation results just mentioned above.  
The host galaxy and standard star observation on 14 September were taken under good condition. Hence we took a relatively small infinite aperture, that is 20 pixels (4 seconds, pixel scale is 0.200 second/pixel). 
The source and the standard star flux were measured using small apertures ($1-2$ times of FWHM) to maximize the ${\rm SNR}$ and finally corrected to infinite apertures. 
The real atmosphere extinction is not important in practice for this case, since the airmass for object and standard star are very close to each other, so the error introduced is less than 0.01 mag. 
The total error is thus the above four uncertainties added in quadrature, as listed in Extended Data Table 1. 
Our $I/R/V/B$ results are similar to that reported in Xu et al.\cite{Xu2009} and the largest deviation is  the dimmer of our $B$ band  emission by $0.2$mag (as shown in Fig.\ref{fig:GRB060505}, our $B$ band flux is naturally in connection with that measured by Gemini in $g'$ band),
which may be benefitted from the improvement of the image subtraction technology in recent years.

{\it HST data processing.}
On 19 May and 6 June, 2006, HST ACS/WFC observed the afterglow of GRB060505 at F475W and F814W band (proposal ID 10551, PI: Shrinivas Kulkarni). 
We downloaded the fully reduced science images from HST Legacy Archive and did the image subtractions. 
We take the Optimal Image Subtraction (OIS; https://github.com/quatrope/ois), which offers 3 methods  to subtract images: 1) Alard-Lupton ($AL$ in short), 2) Bramich, 3) Adaptive Bramich ($AB$ in short). For HST data, the Bramich method gives the best difference images compared among the three. Both difference images of F475W and F814W showed excesses of photon counts at the position of afterglow. The ${\rm SNR}=(4.7,~3.8)$ for F475W and F814W observations, respectively. We adopted apertures ranging from 0.5 to 10 pixels with a step radius of 0.1 pixel to measure photometries of afterglow and got a curve of magnitudes versus aperture radii (M-AR in short). Several unsaturated bright stars were chosen to measure the growth curve of PSF. The growth curve was used to fit data of M-AR ranging from 1 to 5 pixels due to the fact that wings of PSF for faint source are dominated by noise. Then we took the magnitude at  5.5 arcsec (the standard "infinite" aperture for ACS, 110 pixel) of the fitted growth curve as the measured photometry of afterglow. Zeropoints of images, according to ACS Data Handbook Section 5.1.3, can be calculated with their header keywords, and AB zeropoints of F475W and F814W derived with header keywords were 26.065th and 25.955th AB magnitudes, respectively. Such values are consistent with those given by the online ACS Zerpoints Calculator. Our measured fluxes are lower than the upper limits set in Ofek et al.\cite{Ofek2007}, 
as anticipated. Our successful detections are likely due to the improved algorithm of image subtraction. 

{\it UVOT data processing.}
Swfit/UVOT observed afterglow with V, B, U, W1, M2, W2 filters after BAT trigger. 
We started from the level 2 UVOT images, and used the standard UVOT pipeline of HEASoft to stack images. Images taken after 11 May were used as reference images to be subtracted from images taken on 5 May, but there is no proper B band reference image so we do not analyze it. The OIS was applied to subtract images while the AB method gave the best difference images, because the PSF of UVOT is not stable as that of HST. With the standard UVOT pipeline, we used a circle located at the position of afterglow with a radius of 5 arcsec as source region and an annulus with a inner radius of 15 arcsec and an outer radius of 25 arcsec as background region to measure the photometry of afterglow. Only U band image showed excess with a ${\rm SNR}\approx 3.3$, then we measured the curve of magnitudes versus aperture radii to apply the aperture correction to 5 arcsec, like what we had done with HST data. For other bands, 3 sigma upper-limits with an aperture radius of 5 arcsec were given.

{\it Gemini data processing.}
The Gemini South telescope (Gemini-S) observed the field and revisited it on May 12 (under Gemini south proposal GS-2006A-Q-8, PI: Brian Schmidt). Our reduction for Gemini (South) data are essentially the same as VLT data, which include the bias subtraction, flat-field normalization and $i$ band fringing subtraction. 
Image subtraction between the data collected on 6 May and 26 May has revealed the afterglow in both $g'$ and $i'$ bands.
Since there is no corresponding standard star observations for these images, 
we have selected some field star, calibrated their $B$, $V$, $R$, $I$ and $z$ bands magnitude by VLT image and Interpolated them to $g'$ and $i'$ bands, assuming blackbody spectra. 
Then these stars are used to calibrated the afterglow. Our results are almost the same as Ofek et al.\cite{Ofek2007}.

{\bf Swift-XRT data analysis.}
There are in total five observations from the Neil Gehrels {\it Swift} Observatory/XRT 
targeting on the field of GRB 060505 and the first one began at $t\sim$ 13.4 hr after the BAT trigger. 
The XRT photon counting mode data are analyzed by the updated FTOOLS software version 6.28. 
The data reduction procedures include the initial event cleaning carried out with {\tt xrtpipeline} using standard quality cuts. 
A circle region with a radius of 15 pixels is chosen for the extraction of the source spectra, meanwhile,  a larger circle (50 pixels) in a blank area is chosen for the background. To facilitate the sequent spectral analysis, the ancillary response files are made by the {\tt xrtmkarf} task based on the {\tt CALDB} database. 
Among the five observations, X-ray signals towards the target that are distinguishable from the background are seen only in the first two cases (i.e. observation identifiers 00208654001 and 00208654002). 
Due to the limited statistics, the grouped spectra are required to have at least 1 count per bin using the {\tt cstat} approach. We take the Galactic absorption  ($N_{\rm H}$ = 1.8 $\rm \times10^{20}$ $\rm cm^{-2}$, see Ofek et al.\cite{Ofek2007}) and set the intrinsic one from the host galaxy as a free parameter, the first set of {\it Swift} XRT data  yields a hard spectrum of $\propto \nu^{-0.55\pm 0.34}$ (68\% credibility),  in agreement with that reported in {\it UK Swift Science Data Centre} (https://www.swift.ac.uk/xrt$_{-}$spectra/00208654/; where the spectrum $\propto \nu^{-0.5^{+0.7}_{-0.5}}$ is for 90\% credibility)\cite{Evans2009}. 
For the second set of XRT data and the Chandra data, it is not possible to constrain the spectra and we simply adopt the first XRT spectrum. 
In Extended Data Table 2 and Extended Data Fig.1, motivated by the earlier ultraviolet/optical/X-ray data, we take an afterglow spectrum of  
$F_\nu \propto \nu^{-0.6}$ and just estimate these late time fluxes/fluences. 
While Ofek et al.\cite{Ofek2007} assumed a spectrum of either $\nu^{-1}$ or $\nu^{-1.5}$ to estimate the fluxes. Anyhow, the fluxes at $\approx 2$ keV estimated in these two ways are similar. 

{\bf The kilonova emission in the presence of a moderate intrinsic redding.} {In the main text, we have assumed no intrinsic redding in the host galaxy. While in the stellar population modeling at the GRB site the best fit prefers an intrinsic redding of $E(B-V)\approx 0.07$ mag\cite{Thone2008} (In the fit to the stellar population in the host galaxy, $E(B-V)\approx 0.1$mag was favored and the corresponding stellar age is $\approx 10$ Myr\cite{Leibler2010}). An intrinsic $N_{\rm H}\sim 10^{21}~{\rm cm^{-2}}$ is allowed in the X-ray spectral analysis, which may indicate a moderate extinction in the host galaxy\cite{Covino2013}. Therefore, here  we assume $E(B-V)\approx 0.07$ mag  (in the host galaxy) and the extinction law of the Milky Way to examine whether the kilonova properties have been significantly modified. As  shown in Extended Data Fig.4,  though the UVOT upper limits are less tighter than the case of no intrinsic redding, they still play an important role and the optical/X-ray spectrum should be harder than $\nu^{-0.7}$. In the third epoch, the HST and X-ray data are well described by $\propto \nu^{-0.60\pm0.09}$. Therefore,  $\nu^{-0.6}$ is still a reasonable approximation of the forward shock afterglow emission spectrum and a kilonova component is  evident. The intrinsic temperature of the kilonova has been shifted to a higher value of $\sim 6100$ K.}

{\bf The physical origin of the X-ray and optical emission.} As extensively discussed in the main text, the early time optical emission was dominated by a thermal component (i.e., the blue kilonova). At late time, the optical emission was only measured by HST at $t\approx 14$ day. Though it is insufficient to construct a reliable optical afterglow emission lightcurve, such data are indeed essential to construct the  optical to X-ray afterglow spectrum that reads $F_\nu \propto \nu^{-0.6}$ (see Fig.\ref{fig:GRB060505}). This spectrum is typical for GRB afterglow. Within the fireball afterglow model,  such a spectrum can be produced if the typical synchrotron radiation frequency of the shock-accelerated electrons is below the optical band while the cooling frequency of these electrons are above the X-ray band\cite{Piran2004}. Moreover, the power-law distribution index of the shock-accelerated electrons is $p\approx 2.2$. The X-ray afterglow decline should be $\propto t^{-3(p-1)/4} \propto t^{-0.9}$ in the pre-jet break phase and $\propto t^{-p}$ after the jet break. The observed X-ray decline (i.e., $t^{-1.39^{+0.24}_{-0.18}}$) likely point towards a jet break at $t\approx 2.5$ day after the trigger of the GRB. However, in total we have just 3 X-ray data points, which hampers us to reliably infer more information.

{\bf  The kilonova sample.}  Before 2017,  kilonova/macronova signatures had been identified in GRB 130603B\cite{Tanvir2013,Berger2013},  GRB 060614\cite{Yang2015,Jin2015} and GRB 050709\cite{Jin2016}.  After the successful detection of AT2017gfo\cite{Pian2017,Drout2017,Kasliwal2017,Arcavi2017,Cowperthwaite2017}, a few kilonova candidates were reported in GRB 150101B\cite{Troja2018}, GRB 160821B\cite{Jin2018,Troja2019,Lamb2019} and GRB 070809\cite{Jin2020}. 
In Fig.\ref{fig:Comparison} we summarize the evolution of the luminosity, the temperature and the velocity of
the emitting region of the kilonovae. Note that previously $T_{\rm int}$ was only measured for the kilonovae in GRB 060614\cite{Jin2015}, GW170817\cite{Drout2017}, GRB 160821B\cite{Jin2018,Troja2019} and GRB 070809\cite{Jin2020}. For  GRB 130603B and GRB 050709 the upper limits on $T_{\rm int}$ can be set.  The exception is GRB 150101B, which optical emission was only detected in $R$ band and so it is not possible to set any constraint on $T_{\rm int}$. Therefore, for GRB 130603B, GRB 050709 and GRB
150101B, the bolometric luminosities can not be reliably calculated  and following Jin et al.\cite{Jin2020} we simply adopt the approximation of $1.4\nu L_\nu$  (marked in open circles without error bars). It is also not possible to infer 
$\beta_\Gamma$ due to the lack of $T_{\rm int}$ for GRB 130603B, GRB 050709 and GRB 150101B. These data, together with our current GRB 060505, are plotted in Extended Data Fig.2. 
GRB 060505 is similar to other events at the same epoch. Such a similarity provides an additional support to the kilonova origin of the early time optical emission of GRB 060505.

{\bf The time delay distribution sample.} To compare the host galaxy properties of GRB 060505 with others hosting kilonovae, we have collected from their absolute magnitude, star formation rate, stellar population age etc. from literature. The properties are summarized in Extended Data Table 1. 
For GRB 070809, in the case of $z=0.473$, the stellar population age was estimated to be 4.3 Gyr\cite{Leibler2010}. While for the possible host galaxy at $z=0.2185$, following Bolzonella et al.\cite{Bolzonella2000} we estimate the stellar population age  to be $\sim 1$ Gyr with the multi-band photometry data (i.e. Keck $g$, $R$, HST $F606W$ and $F160W$). 
Clearly, GRB 060505 is typical in the host galaxy luminosities but has the shortest delay time. Indeed, a delay time of $6-10$ Myr\cite{Ofek2007,Thone2008,Leibler2010} is at the low end of the expected delay time distribution of the neutron star mergers\cite{Belczynski2006}.

\renewcommand{\refname}{References}


\begin{addendum}

\item[Acknowledgements] This work was supported in part by NSFC under grants of  No. 11933010 and No. 11773078, the Funds for Distinguished Young Scholars of Jiangsu Province (No. BK20180050), the China Manned Space Project (NO.CMS-CSST-2021-A13), Major Science and Technology Project of Qinghai Province  (2019-ZJ-A10), Key Research Program of Frontier Sciences (No. QYZDJ-SSW-SYS024). SC and PD have been supported by ASI grant I/004/11/0.

\item[Author Contributions] Y.Z.F, Z.P.J, S.C, and D.M.W launched the project. Z.P.J, H.Z., N.H.L, L.L, X.L, S.C and P.D carried out the data analysis. Y.Z.F, D.M.W and Z.P.J interpreted the data. Y.Z.F and Z.P.J prepared the paper and all authors joined the discussion.

\item[Author Information]
The authors declare no competing financial interests.
Correspondence and requests for materials should be addressed to Y.Z.F (yzfan@pmo.ac.cn).

\item[Code availability.] The codes used in this analysis are standard in the community, as introduced in the Methods.

\item[Data availability.] The VLT, {HST}, Gemini-S and {\it Swift} observation data analysed/used in this work are all publicly available.

\newpage

\label{tab:comparison}
\title{}{\bf Extended Data Table 1. $\mid$ Comparison of the environments of kilonovae.} \\
\begin{tabular}{lllllllll}\hline\hline
Host	& z	& Type	& $M_{\rm B}$	& log($M$)	& SFR	& Age & offset 	& Ref.	\\
		& 	& 	& mag	& $M_{\odot}$	& $ M_{\odot}~{\rm yr}^{-1}$	& Gyr	& kpc &	\\
\hline
050709	& 0.160	& dwarf 	& -16.9	& 8.66	& 0.2	& 1	& 3.8	& (1-3)\\
060505	& 0.089	& spiral(Sc)	& -19.2	& 9.4	& 1.2	& 0.006-0.01	& 7.1	& (4-7)\\
060614	& 0.125	& dwarf	& -15.5	& 7.95	& 0.013		& 0.57	& 1.1	& (7-10) \\
070809(G1)	& 0.219	& spiral	& -17.7	& 10.15	& 0.15	& 1.0	& 20		&  (11)	\\
070809(G2)	& 0.473	& elliptical	& -21.9	& 11.4	& $<$0.1		& 3.1	& 35		& (7)	\\
130603B	& 0.356	& spiral	& -19.8	& 9.7	& 5.6		& 1.0	& 5.4	& (12)	\\
150101B	& 0.134	& elliptical	& -21.7	& 10.85	& 0.5	& 4	& 7.3	& (13-14)\\
160821B	& 0.161	& spiral 	& -19.9		& 8.5	& 1.2 	& 0.7	& 16.4	& (15-16) \\
170817	& 0.0098	& elliptical/S0	& -20.4	& 10.65	& 0.01	& 4.2	& 2.1	& (17) \\
\hline\hline
\end{tabular}

(1).Fox et al. 2005\cite{sup_Fox2005}; (2).Hjorth et al. 2005\cite{sup_Hjorth2005}; (3).Covino et al. 2006\cite{sup_Covino2005};
(4). Ofek et al. 2007\cite{Ofek2007}; (5). Th\"one et al. 2008\cite{Thone2008}; (6). Th\"one et al. 2014\cite{Thone2014};
(7). Leibler \& Berger 2010\cite{Leibler2010};
(8). Della Valle 2006\cite{sup_Della2006}; (9). Gal-Yam et al. 2006\cite{sup_Gal-Yam2006}; (10). Fynbo et al. 2006\cite{Fynbo2006}; 
(11). Perley et al. 2008\cite{sup_Perley2008} and this work; 
(12). de Ugarte Postigo et al. 2014\cite{sup_deUgartePostigo2014}; 
(13). Troja et al. 2018\cite{Troja2018}; (14). Fong et al. 2016\cite{sup_Fong2016}; 
(15). Troja et al. 2019\cite{Troja2019}; (16). Lamb et al. 2019\cite{Lamb2019};
(17). Fong et al. 2017\cite{sup_Fong2017}.

\renewcommand{\refname}{References}

\newpage

\label{tab:observation}
\title{ \bf Extended Data Table 2. $\mid$ Optical observations of GRB 060505} \\
\begin{tabular}{llllllllll}
\hline \hline
Time  & Exposure & Instrument  & Filter	& Magnitude$^{a}$ \\
(days)    &	(seconds)	& & & (AB)	\\ 
\hline 
\hline
0.70189	& 1666	& {\it Swift}-UVOT	& $UVW1$	& $>22.04$\\
0.70565	& 835	& {\it Swift}-UVOT	& $U$  	& $22.17\pm0.35$\\
0.70817	& 835	& {\it Swift}-UVOT	& $B$	& No proper reference\\
0.71432	& 3335	& {\it Swift}-UVOT	& $UVW2$	& $>23.89$\\
0.72059	& 835	& {\it Swift}-UVOT	& $V$ 	& $>20.80$\\
0.72507	& 2194	& {\it Swift}-UVOT	& $UVM2$	& $>23.47$\\

5.17214	& 3746	& {\it Swift}-UVOT	& $UVW2$	& Reference\\
5.23692	& 4064	& {\it Swift}-UVOT	& $UVM2$	& Reference\\
5.23987	& 2724	& {\it Swift}-UVOT	& $UVW1$	& Reference\\
5.24165	& 1352	& {\it Swift}-UVOT	& $U$  	& Reference\\
11.28599	& 2131	& {\it Swift}-UVOT	& $V$ 	& Reference\\
11.29094	& 2001	& {\it Swift}-UVOT	& $UVW1$	& Reference\\
12.92259	& 1896	& {\it Swift}-UVOT	& $V$ 	& Reference\\
12.92695	& 1736	& {\it Swift}-UVOT	& $UVW1$	& Reference\\
16.53361	& 11608	& {\it Swift}-UVOT	& $V$ 	& Reference\\

1.10191	& 900	& Gemini-S-GMOS	& $r'$	& $21.88\pm0.16$\\
1.11781	& 1200	&  Gemini-S-GMOS	& $g'$	& $22.37\pm0.08$\\

21.01732	& 300	& Gemini-S-GMOS	& $r'$	& Reference\\
21.04284	& 3000	&  Gemini-S-GMOS	& $g'$	& Reference\\

1.11425	& 120	& VLT-FORS1	& z$_{\rm GUNN}$	& $21.47\pm0.08$\\
1.11685	& 80	& VLT-FORS1	& I$_{\rm BESS}$	& $21.50\pm0.04$\\
1.11870	& 80	& VLT-FORS1	& V$_{\rm BESS}$	& $22.10\pm0.04$\\
1.12055	& 150	& VLT-FORS1	& B$_{\rm BESS}$	& $22.59\pm0.04$\\
1.125981	& 734	& VLT-FORS1	& R$_{\rm BESS}$	& $21.83\pm0.04$\\

131.78574	& 300	& VLT-FORS1	& R$_{\rm BESS}$	& Reference\\
131.79030	& 300	& VLT-FORS1	& z$_{\rm GUNN}$	& Reference\\
131.79499	& 300	& VLT-FORS1	& I$_{\rm BESS}$	& Reference\\
131.79939	& 300	& VLT-FORS1	& V$_{\rm BESS}$	& Reference\\
131.80379	& 300	& VLT-FORS1	& B$_{\rm BESS}$	& Reference\\

14.36021	& 7047	& {\it HST}-WFC1	& $F814W$	& $27.77\pm0.30$ \\
14.36160	& 14094	& {\it HST}-WFC1	& $F475W$	& $28.09\pm0.28$ \\

32.68253	& 6840	& {\it HST}-WFC1	& $F814W$	& Reference\\
32.90693	& 13680	& {\it HST}-WFC1	& $F475W$	& Reference\\

\hline

 & & & & Flux\\
 & & & & (erg cm$^{-2}$ s$^{-1}$)\\

\hline

0.698	& 8000	& {\it Swift}-XRT	& 0.3-10keV	& $1.25^{+0.26}_{-0.23}\times10^{-13}$  \\
4.756	& 11700	& {\it Swift}-XRT	& 0.3-10keV	& $1.00^{+0.75}_{-0.55}\times10^{-14}~^{b}$ \\

19.059	& 24710	& {\it Chandra}-ACIS-S	& 0.3-10keV	& $4.71^{+6.28}_{-3.14}\times10^{-15}~^{b}$ \\

\hline \\
\end{tabular}
\\Note: $a$. These magnitudes have been corrected for the Galactic extinctions of $E(B-V)=0.0174$\cite{Schlafly2011}. \\
 $b$. A spectrum of $F_\nu \propto \nu^{-0.6}$ was assumed.\\

\newpage

\begin{figure}[!h]
\begin{center}
\includegraphics[width=0.8 \columnwidth]{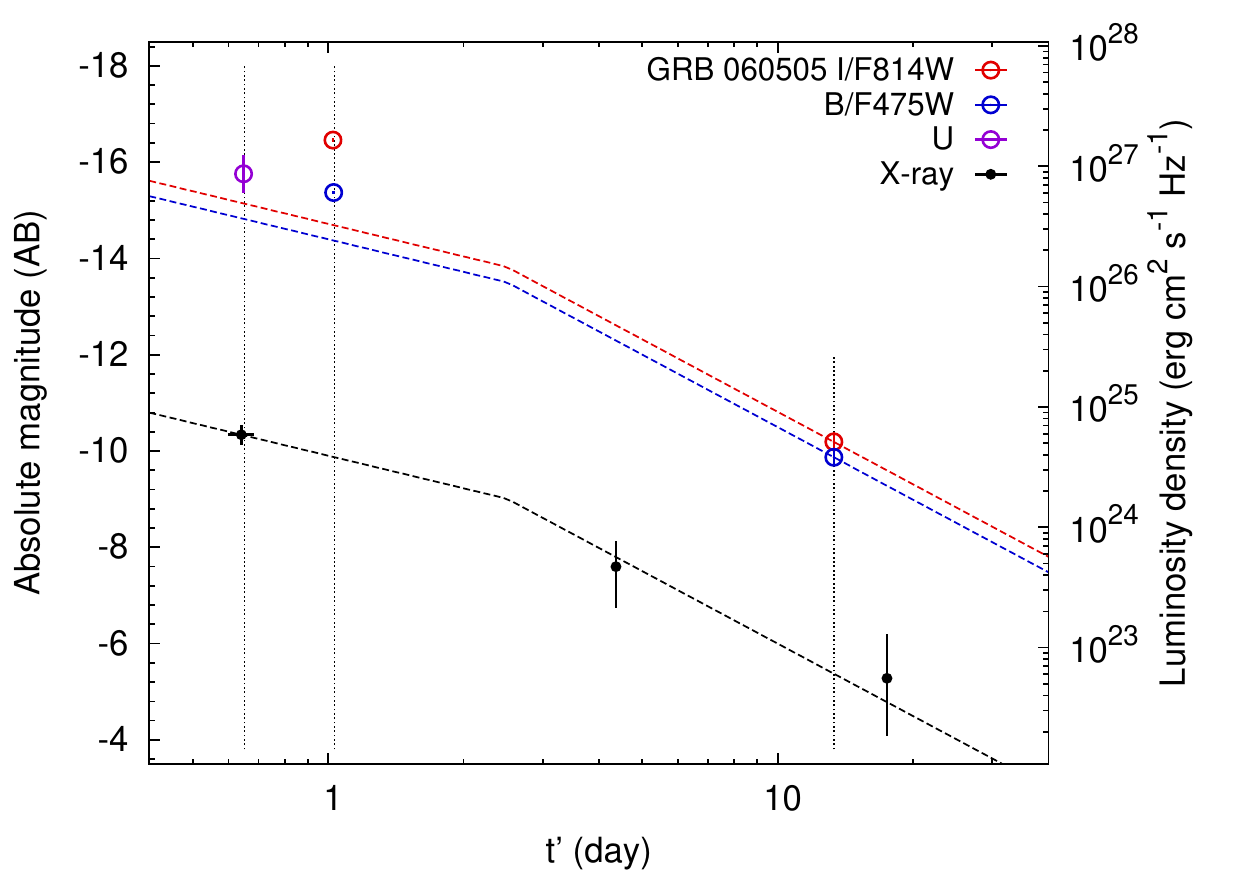}
\end{center}
\label{fig:GRB060505Light}
\end{figure}
{\bf Extended Data Fig.1 $\mid$ Lightcurves of GRB 060505.} 
The X-ray lightcurve is for the frequency of $\approx 2$ keV. The dashed lines represent the forward afterglow component expected in the fireball model (see the Methods). 
Evidently, the early time ultraviolet/optical emission should host an additional component, consistent with that found with the spectra (see Fig.2). 
Vertical dotted lines show the three epochs we adopted for SED fit.

\newpage

\begin{figure}[!h]
\begin{center}
\includegraphics[width=0.75 \columnwidth]{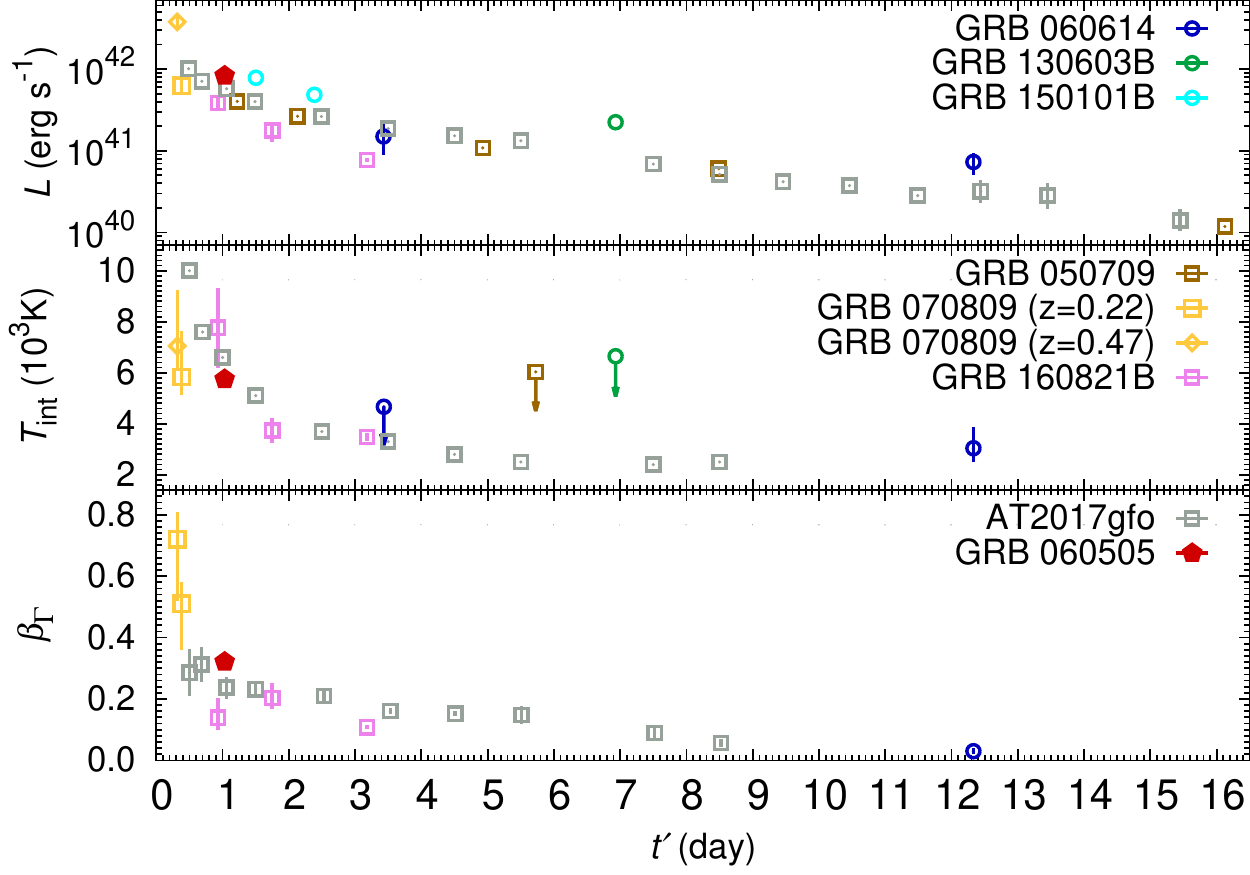}
\end{center}
\label{fig:Comparison}
\end{figure}
{\bf Extended Data Fig.2 $\mid$ Comparison of kilonova properties.} {The data of GRB 060505 is from this work and the rest are from Jin et al.\cite{Jin2020}. In comparison to AT2017gfo, at $t'\approx 1$ day the blue kilonova component of GRB 060505 is more luminous by a factor of $\sim 2$  while the temperature as well as the velocity are comparable.}

\newpage
\begin{figure}[!h]
\begin{center}
\includegraphics[width=0.75\columnwidth]{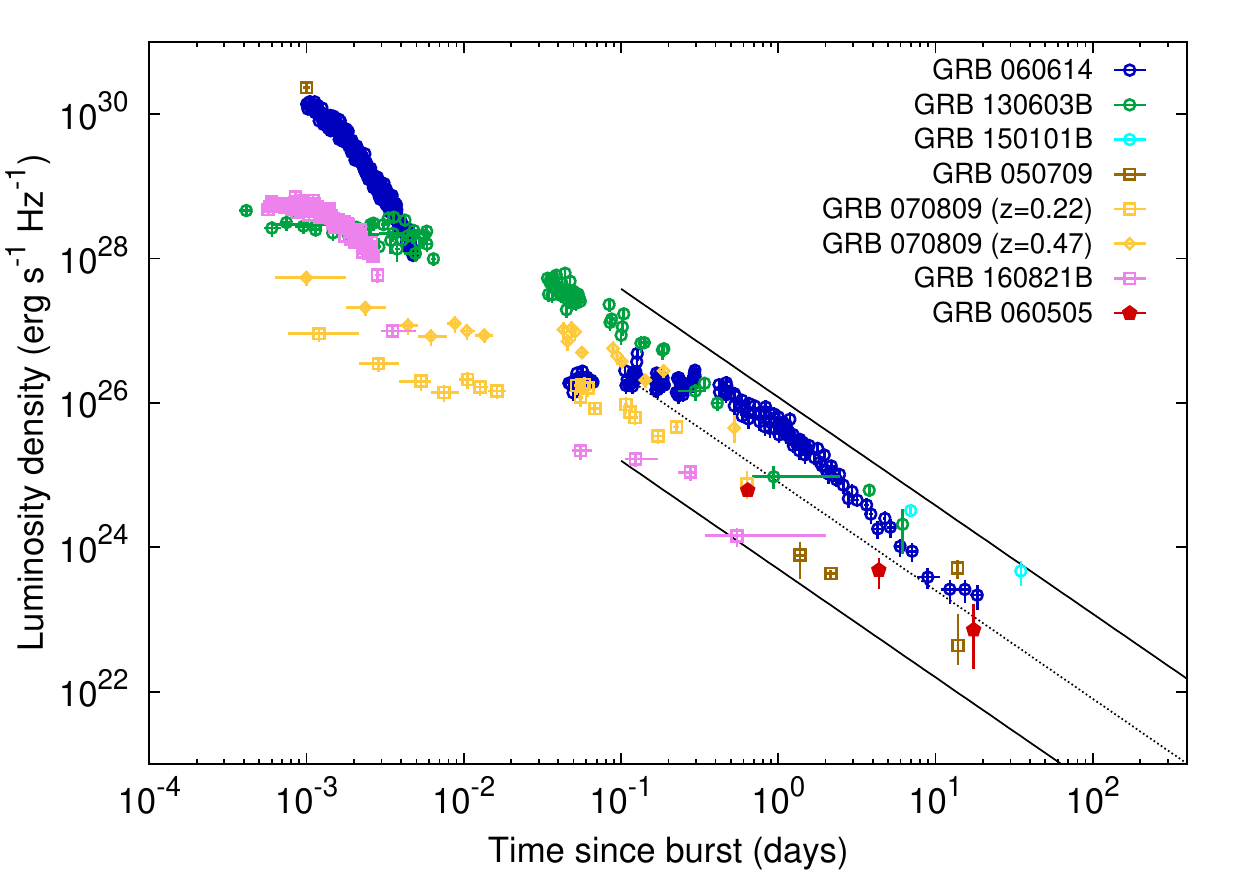}
\end{center}
\label{fig:X-ray}
\end{figure}
{\bf Extended Data Fig.3 $\mid$ Comparison of the X-ray afterglow of GRB 060505 to that of the short/hybrid GRBs displaying kilonova signals.} GW170817/GRB 170817A is excluded  because of its off-axis nature.  In comparison to other events displaying kilonova signal, the X-ray afterglow of GRB 060505 is among the faint group, which renders the identification of the kilonova signal easier. The data of all events except GRB 060505 are adopted from Duan et al.\cite{sup_Duan2019}.

\newpage

\begin{figure}[!h]
\begin{center}
\includegraphics[width=0.8 \columnwidth]{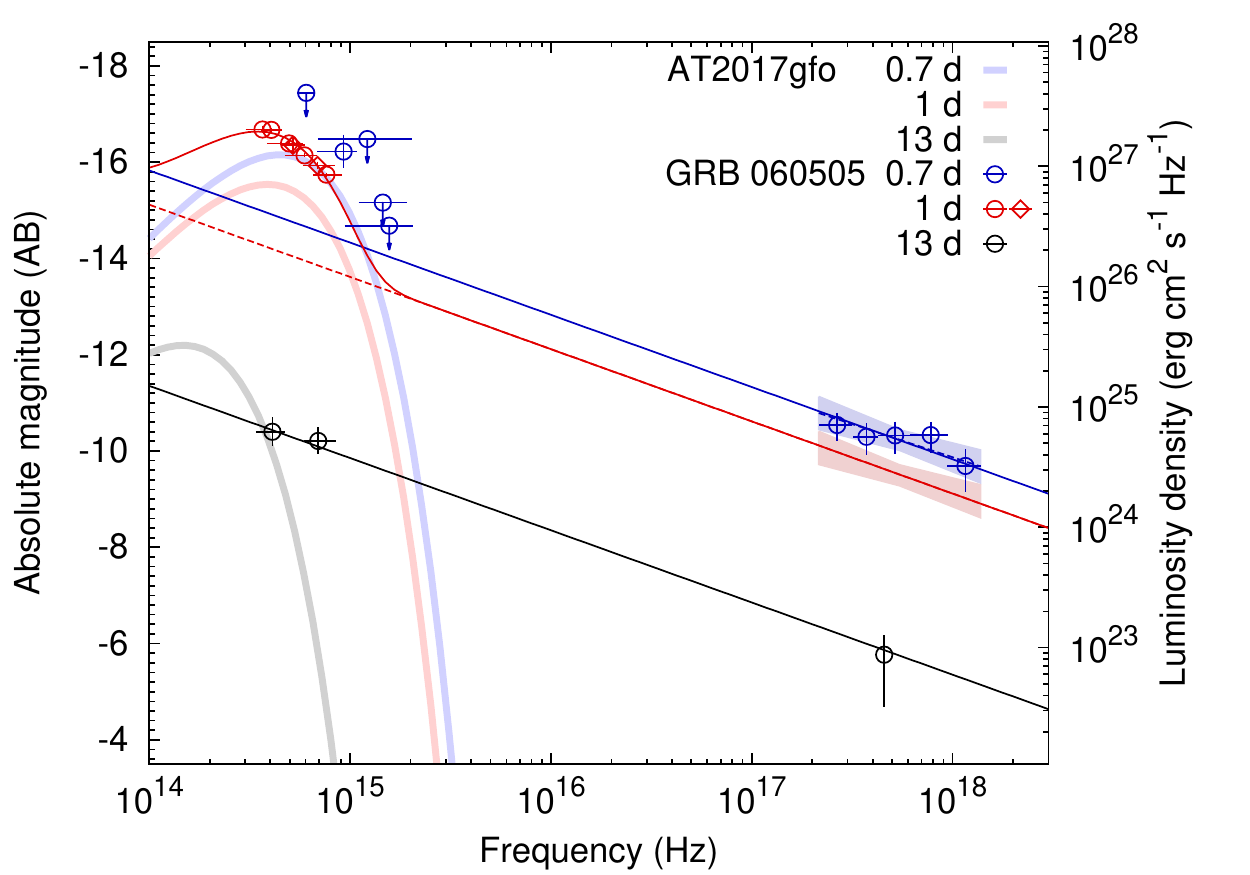}
\end{center}
\end{figure}
{\bf Extended Data Fig.4 $\mid$ Spectral energy distributions  of the ``afterglow" emission of GRB 060505 in the case of ``intrinsic redding".}
 This plot is the same  as Fig.2 except for assuming an additional extinction $E(B-V)=0.07$ mag from the host galaxy. 
The presence of a thermal-like radiation component in the optical/ultraviolet bands is still evident.
\label{fig:GRB060505-SED-Av}
\end{addendum}


\begin{thebibliography}{30}

\bibitem{Fynbo2006} Fynbo, J. P. U. et al. 
{No supernovae associated with two long-duration gamma- ray bursts.}
\newblock  {\it  Nature} {\bf 444}, 1047-1049 (2006).

\bibitem{Ofek2007} Ofek, E. O. et al. 
{GRB 060505: A possible short-duration Gamma-Ray Burst in a star-forming region at a redshift of 0.09.}
\newblock  {\it  Astrophys. J.}  {\bf 662}, 1129-1135 (2007).

\bibitem{McBreen2008} McBreen, S. et al. 
{The spectral lag of GRB 060505: A likely member of the long-duration class.}
\newblock  {\it  Astrophys. J. Lett.}  {\bf 677}, L85-L88 (2007).

\bibitem{Thone2008} Th\"{o}ne, C. C. et al. 
{Spatially resolved properties of the GRB 060505 host: Implications for the nature of the progenitor.}
\newblock  {\it  Astrophys. J.}  {\bf 676}, 1151-1161 (2008).

\bibitem{Thone2014} Th\"{o}ne, C. C. et al. 
{The host of the SN-less GRB 060505 in high resolution.}
\newblock  {\it  Mon. Not. R. Astron. Soc.}  {\bf 676},  2034-2048 (2014).

\bibitem{Piran2004} Piran, T.
The physics of gamma-ray bursts.
\newblock {\it Rev. Mod. Phys.} {\bf 76}, 1143-1210 (2004).

\bibitem{Li1998} Li, L. X. \& Paczy\'{n}ski, B.
Transient Events from Neutron Star Mergers.
\newblock {\it Astrophys. J. Lett.} {\bf 507}, L59-L62 (1998).

\bibitem{Metzger2017} Metzger, B. D.
Kilonovae. \newblock {\it Living Rev. in Relativ.} {\bf 23}, 1 (2019).

\bibitem{Belczynski2006} 
Belczynski, K.,  Perna, R.,  Bulik, T.,  Kalogera, V., Ivanova, N.,  Lamb, D. Q. 
A study of compact object mergers as short Gamma-Ray Burst progenitors.
\newblock {\it Astrophys. J.} {\bf 648},1110-1116 (2006).

\bibitem{Abadie2010} Abbott, B. P. et al.
Prospects for observing and localizing gravitational-wave transients with Advanced LIGO, Advanced Virgo and KAGRA.
\newblock {\it Living Rev. in Relativ.} {\bf 23}, 3 (2020).

\bibitem{Eichler1989} Eichler, D., Livio, M., Piran, T. \& Schramm, D. N.
\ {Nucleosynthesis, neutrino bursts and gamma-rays from coalescing neutron stars.}
\newblock  {\it Nature} {\bf 340}, 126-128 (1989).

\bibitem{Abbott2017} Abbott, T. D. et al.
GW170817: Observation of Gravitational Waves from a Binary Neutron Star Inspiral.
\newblock {\it Phys. Rev. Lett.} {\bf 119}, 161101 (2017).

\bibitem{Goldstein2017} Goldstein, A. et al.
An Ordinary Short Gamma-Ray Burst with Extraordinary Implications: Fermi-GBM Detection of GRB 170817A.
\newblock {\it Astrophys. J. Lett.} {\bf 848}, L14 (2017).

\bibitem{Pian2017} Pian, E. et al.
Spectroscopic identification of r-process nucleosynthesis in a double neutron-star merger.
\newblock {\it Nature} {\bf 551}, 67-70 (2017).

\bibitem{Tanvir2013} Tanvir, N. R. et al.
 A `kilonova' associated with the short-duration gamma-ray burst GRB 130603B.
\newblock {\it Nature} {\bf 500}, 547-549 (2013).

\bibitem{Berger2013} Berger, E., Fong, W. \& Chornock, R.
An r-process Kilonova Associated with the Short-hard GRB 130603B.
\newblock {\it Astrophys. J. Lett.}, {\bf 744}, L23 (2013).

\bibitem{Jin2016} Jin, Z. P. et al.
The macronova in GRB 050709 and the GRB-macronova connection.
\newblock {\it Nat. Commun.} {\bf 7}, 12898 (2016).

\bibitem{Troja2018} Troja, E. et al.
A luminous blue kilonova and an off-axis jet from a compact binary merger at z = 0.1341.
\newblock {\it Nat. Commun.} {\bf 9}, 4089 (2018).

\bibitem{Jin2020} Jin, Z. P., Covino, S., Liao, N. H., et al.
A kilonova associated with GRB 070809.
\newblock {\it Nat. Astron.} {\bf 4}, 77-82 (2020).

\bibitem{Yang2015} Yang, B., Jin, Z. P.,  Li, X., et al.
A possible macronova in the late afterglow of the long-short burst GRB 060614.
\newblock {\it Nat. Commun.} {\bf 6}, 7323 (2015).

\bibitem{Leibler2010} Leibler, C. N. \& Berger, E. 
The Stellar Ages and Masses of Short Gamma-ray Burst Host Galaxies: Investigating the Progenitor Delay Time Distribution and the Role of Mass and Star Formation in the Short Gamma-ray Burst Rate.
{\it Astrophys. J.} {\bf 725}, 1202-1214 (2010).

\bibitem{Palmer2006} Palmer, D. et al.
GRB 060505: Swift-BAT detection of a weak burst. 
{\it GCN Circ.}, 5076, (2006).

\bibitem{Xu2009} Xu, D. et al.
In Search of Progenitors for Supernovaless Gamma-Ray Bursts 060505 and 060614: Re-examination of Their Afterglows
{\it Astrophys. J.} {\bf 696}, 971-979 (2009).

\bibitem{Drout2017} Drout, M. R. et al.
Light curves of the neutron star merger GW170817/SSS17a: Implications for r-process nucleosynthesis.
\newblock {\it Science} {\bf 358}, 1570-1574 (2017)

\bibitem{Kasliwal2017} Kasliwal, M. M. et al.
Illuminating gravitational waves: A concordant picture of photons from a neutron star merger.
\newblock {\it Science} {\bf 358}, 1559-1565 (2017)

\bibitem{Sukhbold2016}
Sukhbold, T., Ertl, T., Woosley, S. E., Brown, J. M., \& Janka, H.-T. 
Core-Collapse Supernovae from 9 to 120 Solar Masses Based on Neutrino-powered Explosions.
\newblock {\it Astrophys. J.} {\bf 821}, 38 (2016).

\bibitem{Yi2006}
Yi, T. F., Liang, E. W., Qin, Y. P., \& Lu, R. J. 
On the spectral lags of the short gamma-ray bursts.
\newblock {\it Mon. Not. R. Astron. Soc.} {\bf 367}, $1751-1756$ (2006).

\bibitem{Abbott2020a} Abbott, R., et al. 
Population Properties of Compact Objects from the Second LIGO-Virgo Gravitational-Wave Transient Catalog.
\newblock {\it Astrophys. J. Lett.} {\bf 913}, L7 (2021).

\bibitem{Kochanek1993} Kochanek, C. S. and  Piran, T.
Gravitational Waves and gamma -Ray Bursts.
\newblock {\it Astrophys. J. Lett.} {\bf 417}, L17 (1993).

\bibitem{LiX2016} Li, X., Hu, Y. M., Fan, Y. Z. \& Wei, D. M.
GRB/GW association: long-short GRB candidates, time-lag, measuring gravitational wave velocity and testing Einstein's equivalence principle.
\newblock {\it Astrophys. J.} {\bf 827}, 75 (2016).




\bibitem{Smith2002} Smith, J. A. et al.
The $u'g'r'i'z'$ standard star system
{\it Astron. J.}, {\bf 123}, 2121-2144 (2002).

\bibitem{Evans2009} Evans, P. A. et al. Methods and results of an automatic analysis of a complete
sample of Swift-XRT observations of GRBs. {\it Mon. Not. R. Astron. Soc.} {\bf 397},
1177–1201 (2009).

\bibitem{Arcavi2017} Arcavi, I. et al.
Optical emission from a kilonova following a gravitational-wave-detected neutron-star merger.
\newblock {\it Nature} {\bf 551}, 64-66 (2017)

\bibitem{Cowperthwaite2017} Cowperthwaite, P. S. et al.
The Electromagnetic Counterpart of the Binary Neutron Star Merger LIGO/Virgo GW170817. II. UV, Optical, and Near-infrared Light Curves and Comparison to Kilonova Models.
\newblock {\it Astrophys. J. Lett.} {\bf 848}, L17 (2017).

\bibitem{Covino2013}  Covino, S., et al. Dust extinctions for an unbiased sample of gamma-ray burst afterglows.
\newblock {\it Mon. Not. R. Astron. Soc.} {\bf 432}, 1231$-$1244 (2013).

\bibitem{Jin2015} Jin, Z. P. et al.
The light curve of the macronova associated with the long-short burst GRB 060614.
\newblock {\it Astrophys. J. Lett.} {\bf 811}, L22 (2015).

\bibitem{Jin2018} Jin, Z. P. et al.
Short GRBs: Opening Angles, Local Neutron Star Merger Rate, and Off-axis Events for GRB/GW Association.
\newblock {\it Astrophys. J.} {\bf 857}, 128 (2018).

\bibitem{Troja2019} Troja, E. et al.
The afterglow and kilonova of the short GRB 160821B. 
 \newblock  {\it  Mon. Not. R. Astron. Soc.}  {\bf 489},  2104-2116 (2019).
 
\bibitem{Lamb2019} Lamb, G. P. et al.
Short GRB 160821B: A Reverse Shock, a Refreshed Shock, and a Well-sampled Kilonova. 
\newblock {\it Astrophys. J.} {\bf 883}, 48 (2019).

\bibitem{Bolzonella2000} Bolzonella, M. et al.
Photometric redshifts based on standard SED fitting procedures.
\newblock {\it Astron. Astrophys.}, {\bf 363}, 476-492 (2000).

\setcounter{firstbib}{\value{enumiv}}

\end{thebibliography}

\begin{thebibliography}{30}

\bibitem[E1]{sup_Fox2005} Fox, D. B., et al.
The afterglow of GRB 050709 and the nature of the short-hard $\gamma$-ray bursts.
\newblock {\it Nature}, {\bf 437}, 845-850 (2005).

\bibitem[E2]{sup_Hjorth2005} Hjorth, J., et al.
The optical afterglow of the short $\gamma$-ray burst
GRB 050709. \newblock {\it Nature}, {\bf 437}, 859-861 (2005).

\bibitem[E3]{sup_Covino2005} Covino, S., et al.
Optical emission from GRB 050709: a short/hard GRB in a star-forming galaxy.
\newblock {\it Astron. Astrophys.}, {\bf 447}, L5-L8 (2006).

\bibitem[E4]{sup_Della2006} M. Della Valle, et al.
An enigmatic long-lasting gamma-ray burst not accompanied by a
bright supernova.
\newblock {\it Nature}, {\bf 444}, 1050 (2006).

\bibitem[E5]{sup_Gal-Yam2006} A. Gal-Yam, et al.
A novel explosive process is required for the $\gamma$-ray burst GRB 060614.
\newblock {\it Nature}, {\bf 444}, 1053 (2006).

\bibitem[E6]{sup_Perley2008} Perley, D. A. et al.
GRB 070809: putative host galaxy and redshift.
{\it GCN Circ.}, 7889, (2008).

\bibitem[E7]{sup_deUgartePostigo2014} de Ugarte Postigo, A. et al.
Spectroscopy of the short-hard GRB130603B: The host galaxy and environment of a compact object merger.
{\it Astron. Astrophy.} {\bf 563}, 62 (2014).

\bibitem[E8]{sup_Fong2016} Fong, W. et al.
The Afterglow and Early-type Host Galaxy of the Short GRB 150101B at $z=0.1343$.
\newblock {\it Astrophys. J.} {\bf 833}, 151 (2016).

\bibitem[E9]{sup_Fong2017} Fong, W. et al. 
The Electromagnetic Counterpart of the Binary Neutron Star Merger LIGO/Virgo GW170817. VIII. A Comparison to Cosmological Short-duration Gamma-Ray Bursts
\newblock {\it  Astrophys. J. Lett.} {\bf 848}, L23 (2019).

\bibitem[E10]{Schlafly2011} Schlafly, E. F. \& Finkbeiner, D. P.
Measuring Reddening with Sloan Digital Sky Survey Stellar Spectra and Recalibrating SFD.
\newblock {\it Astrophys. J.} {\bf 737}, 103–115 (2011).

\bibitem[E11]{sup_Duan2019} Duan, K. K. et al.
Late Afterglow Emission Statistics: A Clear Link between GW170817 and Bright Short Gamma-Ray Bursts. 
\newblock {\it  Astrophys. J. Lett.}  {\bf 876}, L28 (2019).

\end{thebibliography}
\end{document}